# Thermal Hall effect in a van der Waals ferromagnet CrI$_3$

Chunqiang Xu[1,2*], Heda Zhang[1*], Caitlin Carnahan[3], Pengpeng Zhang[1], Di Xiao[4,5], Xianglin Ke[1]

[1]*Department of Physics and Astronomy, Michigan State University, East Lansing, Michigan 48824-2320, USA*

[2]*School of Physical Science and Technology, Ningbo University, Ningbo 315211, China*

[3]*Department of Physics, Carnegie Mellon University, Pittsburgh, Pennsylvania 15213, USA*

[4]*Department of Materials Science and Engineering, University of Washington, Seattle, Washington 98195, USA*

[5]*Department of Physics, University of Washington, Seattle, Washington 98195, USA*

\* These authors contributed equally to this work

CrI$_3$ is a prototypical van der Waals ferromagnet with a magnetic honeycomb lattice. Previous inelastic neutron scattering studies have suggested topological nature of its magnetic excitations with a magnon gap at the Dirac points, which are anticipated to give rise to magnon thermal Hall effect. Here we report thermal transport properties of CrI$_3$ and show that the long-sought thermal Hall signal anticipated for topological magnons is fairly small. In contrast, we find that CrI$_3$ exhibits an appreciable anomalous thermal Hall signal at lower temperature which may arise from magnon-phonon hybridization or magnon-phonon scattering. These findings are anticipated to stimulate further neutron scattering studies on CrI$_3$ single crystal, which can shed light not only on the intrinsic nature of magnetic excitations but also on the magnon-phonon interaction.

# I. INTRODUCTION

The discovery of exotic electronic properties in monolayer graphene has opened a new research arena of searching for novel quasi-two-dimensional (2D) van der Waals (vdW) materials beyond graphene [1, 2]. One prominent class of such materials are vdW magnets. With magnetism acting as a control knob, 2D vdW magnets provide an exciting playground for the discovery and exploration of novel physical phenomena and perhaps new physics [3-6]. In addition to various magnetic ground states and unconventional magnetism, vdW magnets also display many other novel phenomena. For instance, ferromagnetism can persist down to the monolayer limit owing to the material's magnetic anisotropy [7, 8]. Depending on the layer numbers, the magnetic ground state of exfoliated vdW magnets varies [7]. One can also tune the magnetic ground properties using various non-thermal perturbations, such as strain [9, 10], electric gating [11, 12], proximity effect [13-16], etc. Furthermore, new phenomena – so-called Moiré magnetism – can emerge by stacking layers of vdW magnets on top of each other with certain twisting angles [17-19].

In addition to the interesting static magnetic properties, another intriguing aspect of vdW magnets is due to the potential topological characteristic of their magnetic excitations. Particularly, vdW ferromagnets with magnetic ions occupying a Kagome lattice or a honeycomb lattice were proposed to host topological magnons [20, 21]. In a ferromagnetic Kagome lattice or honeycomb lattice, there are six symmetry-protected magnon band crossings at K-points, which retain their degeneracy in the presence of symmetric exchange interactions (e.g. Heisenberg interactions). Because of the inversion symmetry breaking relative to the center of the nearest-neighbor bond of Kagome lattice or the second-nearest neighbor bond of honeycomb lattice, the asymmetric Dzyaloshinskii-Moriya interaction (DMI) consequently lifts the degeneracy at these K-points and leads to the opening of magnon gaps. By calculating the Berry curvature of these gapped magnon bands and the associated Chern number, it has been shown that the gapped magnon bands are topological in nature [20, 21]. Such unique magnon modes can give rise to thermal Hall effect (THE), i.e., an induced transverse heat flow in the presence of a longitudinal temperature gradient [22, 23].

The magnon gap opening at K-points and the emergence of THE have been reported in Cu(1,3–bdc) which has a Kagome lattice of $Cu^{2+}$ ions, suggesting that the magnetic excitations of Cu(1,3-bdc) are topological in nature [24, 25]. For a ferromagnetic honeycomb lattice, the most salient system is $CrI_3$ in which the inelastic neutron scattering (INS) studies have found a gap opening at the K-points of magnon bands [26]. However, the underlying mechanism of the observed gap opening remains controversial [26-31]. For instance, is it driven intrinsically by the DMI [26] or the Kitaev interaction [28], or associated with the interlayer couplings [30], or perhaps driven extrinsically by the mosaic of co-aligned samples for neutron scattering studies [27]? Surprisingly, to date there is no report of THE in literature on $CrI_3$, despite the fact that the magnon THE provides smoking-gun evidence of the topological character of low-energy magnetic excitations.

In this work, we report the thermal transport studies of $CrI_3$ single crystals. we find that the thermal Hall signal associated with the anticipated topological magnetic excitations is fairly small, about an order in magnitude smaller than the predicted value based on the reported DMI values in previous INS studies [26, 27]. In contrast, we show that $CrI_3$ exhibits an appreciable anomalous thermal Hall signal at low temperature, which is presumably of phononic origin. Potential mechanisms of the THE features observed in $CrI_3$ are discussed.

## II. EXPERIMETNAL METHODS

CrI$_3$ has a layered structure with a honeycomb layer of chromium atoms separated by two layers of iodine atoms along the $c$-axis, as shown in Fig. S1 in the Supplemental Material [32]. It undergoes a paramagnetic-to-ferromagnetic phase transition with $T_c$ ~ 60 K and spins aligned along the $c$-axis [33]. In this work, single crystals of CrI$_3$ were synthesized using chemical vapor transport method [33]. The magnetic susceptibility data were measured using a Superconducting Quantum Interference Device (SQUID, Quantum Design). The thermal transport measurements were conducted using one-heater three-sensor technique on a sample puck adapted to a Physical Property Measurement System (PPMS, Quantum Design). The longitudinal and transverse thermal conductivity were simultaneously measured using a conventional steady-state method. The magnetic field was applied along the magnetic easy axis, i.e., $c$-axis. A schematic of the experimental setup for thermal transport measurements is shown in Fig. S2 [32], and more detailed information can be found in the Supplemental Material of Ref. [34]. To minimize exposing the CrI$_3$ samples to the air (moisture), the devices were prepared inside a glove bag equipped with an internal microscope and a continuous flow of high purity nitrogen gas.

### III. RESULTS AND DISCUSSION

Figure 1(a) shows the temperature dependence of magnetic susceptibility ($\chi_c$) of CrI$_3$ measured under zero-field-cooled (ZFC) condition with an applied field of 0.1 T $(\mu_0\vec{H} \parallel \vec{c})$. $\chi_c$ sharply increases below the ordering temperature $T_c$ of ~ 60 K and then saturates, indicative of a ferromagnetic phase transition, which is consistent with the literature [33]. In Fig. 1(b), we present the thermal conductivity ($\kappa_{xx}$) measured as a function of temperature at 0 T and 5 T magnetic field $(\mu_0\vec{H} \parallel \vec{c})$. There are a couple of noteworthy features. First, no anomaly in $\kappa_{xx}$ is obviously observed near $T_c$. Since CrI$_3$ is an insulator, its total $\kappa_{xx}$ is composed of magnon ($\kappa^{mag}$) and phonon ($\kappa^{ph}$) contributions. Thus, the little effect of magnetic ordering on $\kappa_{xx}$ suggests that $\kappa^{ph}$ dominates over $\kappa^{mag}$ in CrI3, a feature distinct from its sister compound CrCl3 [35]. Second, the change in $\kappa_{xx}$ induced by an applied magnetic field ($\Delta\kappa_{xx} = \kappa_{xx}^{5\,T} - \kappa_{xx}^{0\,T}$) is very subtle compared to the signal itself. Therefore, to better quantify this change, we plot $\Delta\kappa_{xx}$ as a function of temperature in Fig. 1(c) together with the scaled temperature derivative of $\frac{1}{\chi_c}\left|\frac{d\chi_c}{dT}\right|$ for reference. We can see that both curves exhibit an upward trend above $T_c$, which is due to the onset of short-range magnetic correlations prior to the long-range order. The positive value of $\Delta\kappa_{xx}$ in the high-temperature regime indicates the existence of magnon-phonon coupling in CrI3. While magnons and phonons contribute individually to the thermal conduction, they also scatter each other, which tends to suppress the total thermal conduction. At high temperatures, in the absence of magnetic field magnons (or incoherent magnons right above $T_c$) scatters phonons, which reduces the $\kappa^{ph}$ contribution and dominates over the $\kappa^{mag}$ term. An applied magnetic field suppresses magnon population by lifting the magnon bands to higher energy, which subsequently reduces the phonon scattering by magnons and (partially) restores the $\kappa^{ph}$ contribution and thus enhances the total $\kappa_{xx}$, leading to positive $\Delta\kappa_{xx}$. Similar feature has been observed in another CrI3 sample as shown in Fig. 1 and in Fig. S3 [32] and in other vdW magnets, such as CrCl3 [35], VI3 [34], and FeCl2 [36]. $\Delta\kappa_{xx}$ changes the sign below 6 K, implying that in the lower temperature region the field-induced reduction of magnon contributions to $\kappa_{xx}$ is larger than the restoration of phonon contribution due to the suppression of phonon scattering by magnons. Note that while the low-temperature $\kappa_{xx}$ value of CrI3 is nearly an order in magnitude larger than that of VI3 (Fig. S4), $\Delta\kappa_{xx}/\kappa_{xx}$ is comparable which implies a similar strength of magnon-phonon coupling in both compounds.

We move on to discuss the transverse thermal signal of CrI3. As will be shown later, it is worth noting that the magnitude of the intrinsic thermal Hall resistivity $|w_{yx}|$ of CrI3 (in the order of $10^{-5}$ $K\,m\,W^{-1}$) is found to be much smaller than that of the longitudinal thermal resistivity $w_{xx}$. Therefore, when dealing with signals with such small magnitude, validating the intrinsic nature of the observed signal is pressing [37]. Here, we examine the nature of the THE observed in CrI3 by studying the hysteresis behavior, the applied heating power dependence, and the sample dependence. We first present the hysteresis behavior of both longitudinal and transverse thermal resistivity and demonstrate that these two measured quantities exhibit distinct behaviors. In Fig. 2(a-b), we present the raw measurement data of $w_{xx}$ and $w_{yx}$ measured at $T$ = 23.7 K. Here the orange and blue curves represent the data measured during field ramping down (1 T to -1 T) and ramping up (-1 T to 1 T) processes, respectively. In contrast to the measured $w_{xx}$ which is nearly symmetric between positive and negative fields [Fig. 2(a)], we can see that $w_{yx}$ shows an asymmetry between positive and negative fields that is superimposed with a large symmetric (even) component. Such an even component stems from the $w_{xx}$ contribution introduced due to a small misalignment between the transverse temperature leads, since the intrinsic $|w_{yx}|$ is much smaller than $w_{xx}$. Generally, one performs anti-symmetrization to eliminate the longitudinal contribution to the measured $w_{yx}$. In Fig. 2(c-d) we present the odd component [$w^{odd} = (w^{H+} - w^{H-})/2$] of both longitudinal and transverse thermal resistivity as $w_{xx}^{odd}$ and $-w_{yx}^{odd}$, where $w^{H+}$ ($w^{H-}$) stands for the thermal resistivity measured in the positive (negative) field region. One can see that while the $w_{yx}^{odd}$ data obtained for these two processes are similar to each other, $w_{xx}^{odd}$ shows an opposite trend for two field ramping processes, which arises from the hysteresis of $w_{xx}$. This suggests that a small misalignment between the transverse temperature leads can cause $w_{yx}^{odd}$ to pick up an extrinsic contribution from $w_{xx}^{odd}$ which cannot be eliminated by simply anti-symmetrizing the raw $w_{yx}$ data. Instead, this extrinsic contribution $w_{xx}^{odd}$ can be properly eliminated by averaging the $w^{odd}$ data extracted from both field ramping processes. The resultant asymmetric components [$w^{Ass} = (w^{odd,up} + w^{odd,down})/2$] for the longitudinal and transverse thermal signal are shown in Fig. 2(e-f). We can see that $w_{xx}^{Ass}$ shown in Fig. 2(e) is nearly zero. In contrast, the thus-extracted $w_{yx}^{Ass}$, the red symbols shown in Fig. 2(f), shows an odd function of magnetic field whose general shape is similar to the magnetization curve (black dash line) measured at the same temperature, representing an intrinsic transverse thermal Hall signal. Similar data processing procedures have been previously applied to extract the intrinsic thermal Hall signal in other systems, such as α-RuCl$_3$ [37] and FeCl$_2$ [36].

We also study the dependence of $w_{yx}^{Ass}$ on the heating power $P$. For an intrinsic THE, the transverse temperature difference $\Delta T_{yx}$ scales linearly with $P$, as implied from the definition of $w_{yx}$ $\left(w_{yx} = \frac{\Delta T_{yx}}{P} t\right)$ where $t$ is thickness of the sample. As shown in Fig. S5, at three different applied heating powers (800 $\mu$W, 1000 $\mu$W, 1470 $\mu$W), the $w_{yx}^{Ass}$ signal is nearly identical, indicating that the thermal Hall signal is indeed independent of $P$. We have also reproduced similar results on another piece of CrI$_3$ sample [See Fig.S6]. These results affirm the intrinsic nature of the THE observed in CrI$_3$.

Figure 3 presents the magnetic field dependence of thermal Hall conductivity $(\kappa_{yx})$, which is defined as $\left(\kappa_{yx} = \frac{-w_{yx}^{Ass}}{w_{xx}^2 + w_{yx}^{Ass^2}}\right)$, measured at various temperatures. The dashed curves are guides to eyes. At low field $\kappa_{yx}$ increases sharply alike the $M(H)$ curve seen in Fig. 2(f), suggesting that at low field $\kappa_{yx}$ mainly arises from the anomalous THE. Upon increasing the magnetic field, there is an additional $\kappa_{yx}$ component that is linearly proportional to the magnetic field. As shown in Fig. S7, the temperature dependence of the coefficient of this linear THE component displays a broad peak near the same temperature as in the $\kappa_{xx}(T)$ curve. These features are similar to those observed in paramagnetic SrTiO$_3$ [38] or antiferromagnetic Cu$_3$TeO$_6$ [39], implying that this linear THE component is purely driven phonons the mechanism of which is beyond the scope of current study.

We now focus on the anomalous THE component. To better quantify the temperature dependence of the anomalous thermal Hall signal, Figure 4(a) shows the temperature dependence of the extracted anomalous thermal Hall conductivity $\kappa_{yx}^A$ (-$\kappa_{xy}^A$). There are two features worth pointing out. First, $\kappa_{yx}^A$ exhibits a broad peak around 10 K, which is close to the temperature at which the longitudinal thermal conductivity $\kappa_{xx}$ maximizes. Similar $\kappa_{yx}^A(T)$ behavior is also observed on another sample (Sample 2) shown in Fig. 4(a), affirming its reproducibility. This suggests that the observed $\kappa_{yx}^A$ in this temperature region is related to phonon. However, here $\kappa_{yx}^A$ could not be purely driven by phonon, as evidenced by the extracted $\kappa_{yx}^A(H)$ shown in the inset of Fig. 4(a) which resembles the *M(H)* curve shown in Fig. 2(f). Instead, phonons are responsible for the anomalous THE component $\kappa_{yx}^A$ via magnon(spin)-phonon coupling due to the magnetoelastic interaction. Similar features have been observed in the VI$_3$ system [34]. A plausible mechanism is associated with magnon-phonon hybridization. Magnons and phonons hybridize at the band anticrossing points, forming a new type of quasiparticle, the magnon-polaron. Berry curvature hotspots are induced in the region near the anticrossing points between the two hybridized bands, which are anticipated to give rise to a THE [40-43]. Recent first-principles calculations showed that, without considering the spin-lattice coupling, the low-energy magnon bands and the acoustic phonon bands cross each other at ~ 1.5 meV and 5.0 meV, at which an observation of the hybridization gap is anticipated when spin-lattice coupling is taken into account [31]. A future INS study to directly probe potential magnon-phonon hybridizations in CrI$_3$ is desirable. Another plausible mechanism to explain the anomalous THE component is magnon-phonon scattering. As discussed previously, phonons are scattered by magnons, which results in slight suppression of $\kappa_{xx}$. It was recently predicted that, similar to the extrinsic anomalous Hall effect induced by skew scattering and side-jump scattering of charge carriers, scattering of phonons by collective fluctuations may give rise to the THE [44].

Second, the $\kappa_{yx}^A$ value drops quickly to nearly zero in the high temperature region. This finding is a surprise and it is in sharp contrast to the $\kappa_{yx}^A(T)$ observed in VI$_3$ in which a large thermal Hall signal is observed over a wide range of temperature nearly up to $T_c$ [34]. As discussed in the introduction, the sizable magnon gap at K points observed in CrI$_3$ via INS studies, which was ascribed to the DMI [26, 27], suggests the topological nature of magnetic excitations in CrI$_3$ that is anticipated to lead to the THE. To support this, we have simulated the magnon bands using a Hamiltonian consisting of exchange parameters extracted from the most recent INS studies [27] and calculated the corresponding thermal Hall conductivity. As shown in Fig. 4(b), with a DMI of 0.09 meV that was recently reported [27], an appreciable thermal Hall signal is anticipated in the high temperature region with a maximum value of ~ 0.01 Wm$^{-1}$K$^{-1}$, which is nearly an order in magnitude larger than the experimental observation presented in Fig. 4(a). This raises an intriguing question: Why is the observed thermal Hall signal in CrI$_3$ so small in the high temperature region compared to the theoretical prediction?

One possible scenario is that the DMI value extracted by fitting the magnetic excitations obtained based on the INS studies using the linear spin wave theory might be overestimated. Indeed, a recent INS study found that the mosaic of co-aligned CrI$_3$ single crystals artificially contributes to the magnons gap opening at K points, which can lead to an overestimation of the DMI value [26, 27]. This is also strongly supported by recent INS studies on a single, large piece of single crystal of CrBr$_3$ [45] and CrCl$_3$ [46] in which the magnon bands cross each other at K points within the instrumental energy resolution, which is in contrast to the observation of a large magnon gap reported in co-aligned CrBr$_3$ single crystals [47]. These results suggest that the intrinsic DMI is neglectable in both CrBr$_3$ and CrCl$_3$ [45, 46]. Although the spin-orbit coupling of iodine atoms is larger than that of bromine and chlorine atoms, which may lead to a larger DMI in CrI$_3$ than that of CrBr$_3$ and CrCl$_3$, it is likely that the previously determined DMI in CrI$_3$ based on the most recent INS studies is still overestimated due to the 8° sample mosaic of the co-aligned single crystals (compared to 17° reported in an earlier study) [26, 27]. In Fig. 4, we plot the calculated thermal Hall conductivity based on a 2D monolayer model with various DMI values. We can see that the calculated thermal Hall conductivity with a DMI of 0.01 meV better agrees with the experimental data.

Another possible mechanism to account for the small thermal Hall signal may be ascribed to the combined effects of the interlayer coupling and the third-neighbor intralayer interaction in CrI$_3$. Note that the sign of $\kappa_{xy}^A$ in CrI$_3$ is opposite to that found in VI$_3$ [34]. Recent theoretical calculations showed that the thermal Hall conductivity in CrI$_3$ strongly depends on both the interlayer coupling and the 3$^{rd}$ neighbor intralayer coupling J$_3$ and that even the sign of $\kappa_{xy}$ can change [29]. With certain phase parameters, the magnetic excitation becomes gapless Weyl type instead of topological magnon insulator, which gives rise to zero/small thermal Hall signal [29]. INS studies on a large, single piece of CrI$_3$ single crystal are demanding, which may allow further determination of the exchange interactions and DMI without the need to consider the extrinsic effect stemming from the mosaic of co-aligned samples. This will help to clarify the topological nature of magnetic excitations in CrI$_3$ and to clarify the mechanisms that account for the thermal Hall signal observed in this study.

Finally, we would like to briefly comment the potential effects of magnon-magnon interactions on $\kappa_{xx}$. Due to the absence of particle-number conservation law for bosonic excitations, the magnon-magnon interactions can affect the topology of magnetic excitations and thus affect the THE. On the one hand, magnon-magnon interactions can give rise to magnon damping, which is anticipated to suppress $\kappa_{xy}$ [48]. However, magnon damping in CrI$_3$ was not discernable in the previous INS studies [26, 27]. It is likely that the higher-order magnon-magnon interactions need to be taken into account for the suppression of $\kappa_{xy}$, as argued in a recent study of SrCu$_2$(BO$_3$)$_2$ [49]. On the other hand, a recent theoretical study suggested that, instead of being detrimental to topology, magnon-magnon interactions can serve as a source of topology of magnon bands and lead to the THE [50].

## IV. CONCLUSION

In summary, we report thermal transport properties of the 2D vdW honeycomb ferromagnet CrI$_3$, a long-sought topological magnon insulator candidate. We show that the thermal Hall signal in the higher temperature region, which is anticipated for topological magnon insulators, is fairly small. An intrinsically small DMI or the cooperative effect of interlayer coupling and intralayer interaction may account for this observation. In contrast, we find that CrI$_3$ exhibits THE with an appreciable anomalous thermal Hall signal in the lower temperature region which may arise from magnon-phonon hybridization or magnon-phonon scattering. These findings will stimulate future INS studies on a large, single piece of CrI$_3$ crystal, which can shed light not only on the magnon-phonon interaction but also on the intrinsic nature of magnetic excitations, which will in turn help elucidate the observed THE phenomena in this study.

**Acknowledgements**

H.Z. and X.K. acknowledge the financial support by the U.S. Department of Energy, Office of Science, Office of Basic Energy Sciences, Materials Sciences and Engineering Division under DE-SC0019259. X.K. also acknowledges the financial support by the National Science Foundation (DMR-2219046). P.Z. acknowledges the financial support by the National Science Foundation (DMR-2112691). C.C. and D.X. are supported by AFOSR MURI 2D MAGIC (FA9550-19-1-0390). C.X. is partially supported by the Start-up funds at Michigan State University.

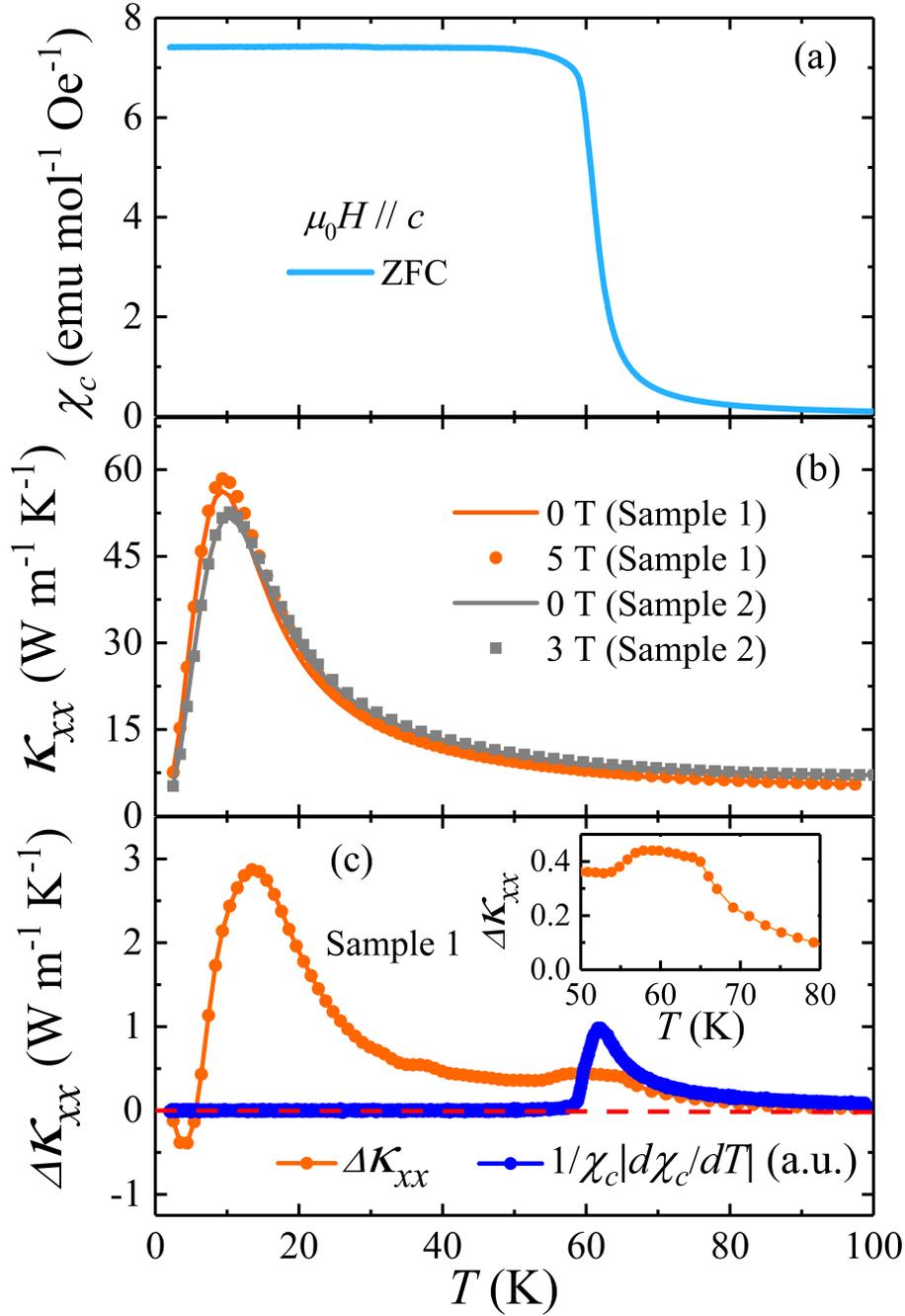

Figure 1: (a) The temperature dependence of magnetic susceptibility ($\chi_c$) measured at 0.1 T under zero-field-cooling (ZFC) condition. (b) The temperature dependence of $\kappa_{xx}$ of two samples measured at 0 T and high magnetic fields. (c) Temperature vs $\Delta\kappa_{xx}$ ($\Delta\kappa_{xx} = \kappa_{xx}|_{5\,T} - \kappa_{xx}|_{0\,T}$) (black) and temperature derivative of magnetic susceptibility $\frac{1}{\chi_c}\left|\frac{d\chi_c}{dT}\right|$ (red).

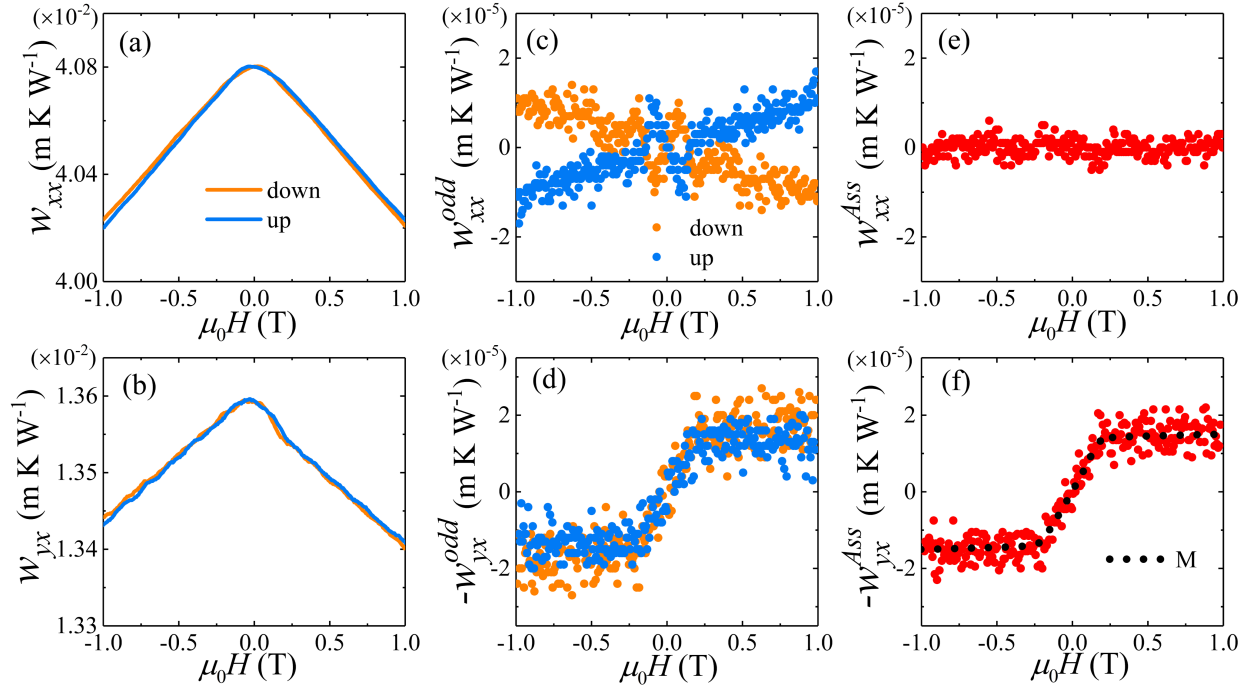

Figure 2: (a) Thermal resistivity $w_{xx}$ hysteresis between -1 T and 1 T. (b) Thermal hall resistivity $w_{yx}$ hysteresis between -1 T and 1 T. (c-d) The odd component $w_{xx}^{odd}$ and $-w_{yx}^{odd}$ ($w^{odd} = (w^{H+} - w^{H-})/2$), which are from (a-b). (e-f) Asymmetric component $w_{xx}^{Ass}$ and $-w_{yx}^{Ass}$ ($w^{Ass} = (w^{odd,up} + w^{odd,down})/2$), which are from (c-d), the dash black line is magnetization curve $M(H)$. Data were measured at $T$ = 23.7 K.

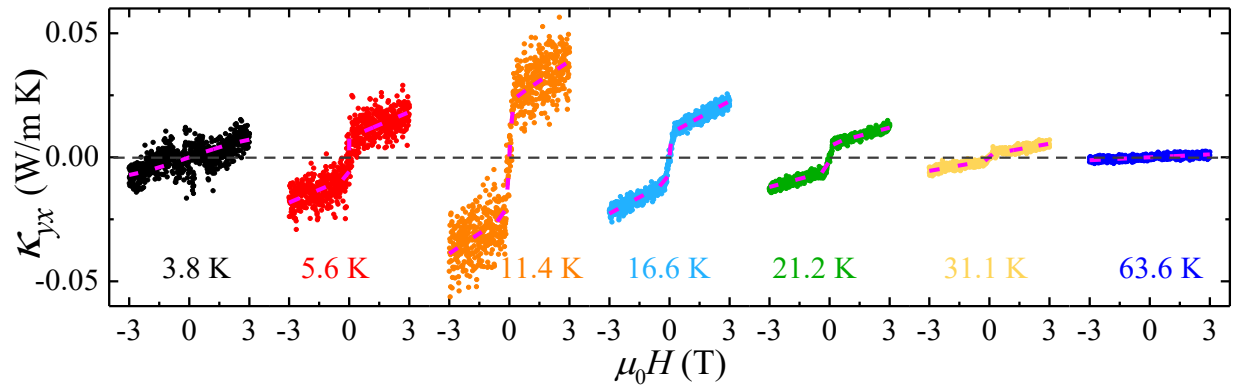

Figure 3: The thermal Hall conductivity $\kappa_{yx}$ measured at various temperatures.

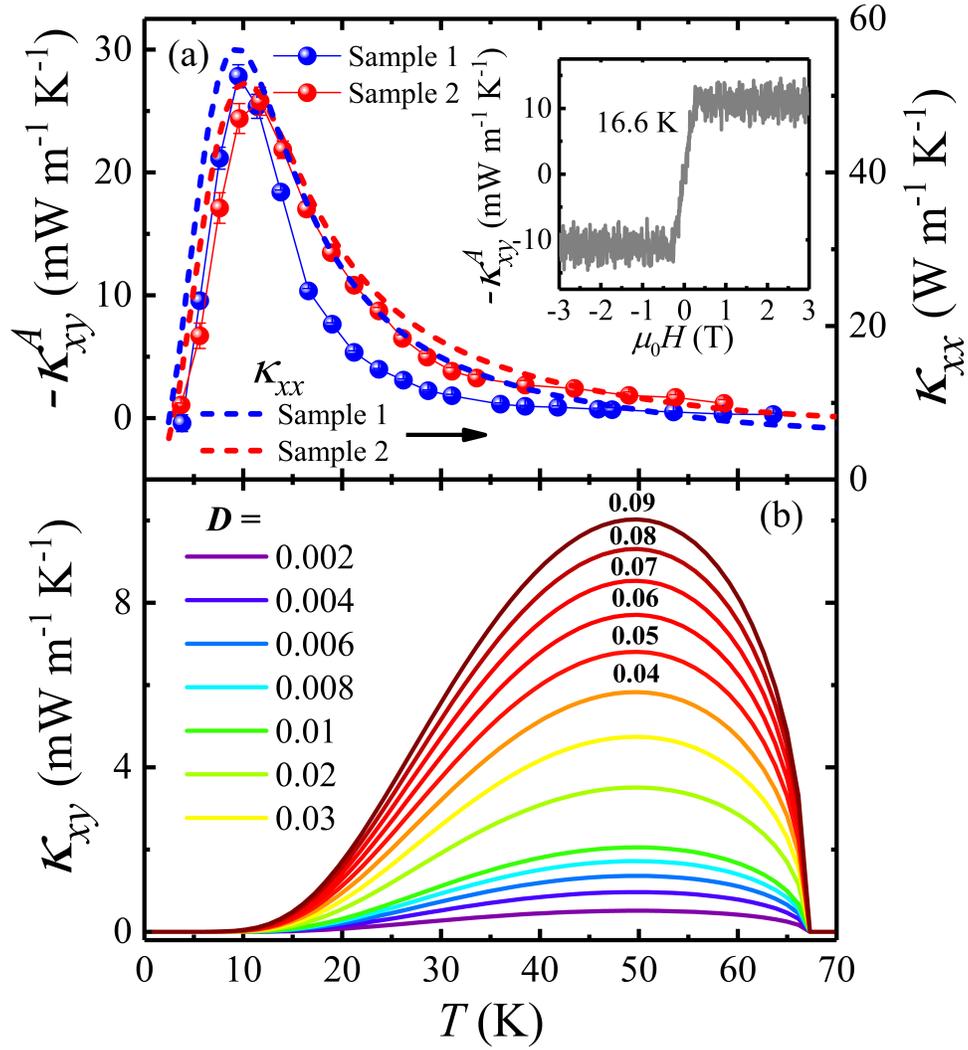

Figure 4: (a) Temperature dependence of anomalous thermal Hall conductivity $-\kappa_{xy}^A$. The $\kappa_{xx}(T)$ curves are overplotted for reference. Inset shows the extracted $-\kappa_{xy}^A$ as a function of field at 16.6 K. (b) The calculated $\kappa_{xy}$ with various DMI values ($D$).